# The Effect of High-Pressure Torsion on Irradiation Hardening of Eurofer-97

Gregory Strangward-Pryce[1, *], Kay Song[2], Kenichiro Mizohata[3], Felix Hofmann[2, &]

[1] Department of Materials, University of Oxford, Parks Road, Oxford, OX1 3PH, UK

[2] Department of Engineering Science, University of Oxford, Parks Road, Oxford, OX1 3PJ, UK

[3] University of Helsinki, P.O. Box 64, 00560 Helsinki, Finland

[*] *gregory.strangward-pryce@ccc.ox.ac.uk*

[&] *felix.hofmann@eng.ox.ac.uk*

## Highlights

- Strain hardness saturation is approached at shear strains > 70 in Eurofer-97.
- Irradiation hardening was less in HPT-deformed samples than in undeformed material.
- Nano-structuring provides a way to reduce irradiation-induced change in Eurofer-97.

## Abstract

We investigated the effect of nano-structuring by high-pressure torsion (HPT) on the irradiation performance of Eurofer-97. Material was deformed to shear strains from 0 to ~230, and then exposed to $Fe^{3+}$ irradiation doses of 0.01 and 0.1 displacements-per-atom (dpa). Nanoindentation hardness increases monotonically with deformation, and with irradiation for the undeformed material. For both damage levels, less irradiation hardening is observed in severely deformed material. This effect is most prominent in the strain range ~60 to ~160, suggesting that nano-structuring may provide an approach for reducing irradiation hardening.

## Keywords

HPT, Irradiation, Eurofer-97, Nanoindentation, Hardening.

1. <u>**Introduction**</u>

Eurofer-97 is proposed for structural components in DEMO [1,2], the steppingstone to commercial fusion power [3]. Eurofer-97, a reduced-activation ferritic/martensitic steel (RAFM) [4], offers good mechanical properties, low long-term activation, and good resistance to irradiation creep and swelling [1]. In service, it will be exposed to intense neutron bombardment and a wide range of temperatures [5]. To increase the in-service life of Eurofer-97 components, a high irradiation resistance must be achieved.

Ion-irradiation can be used to mimic high-flux neutron bombardment of RAFM steels [4,6]. It causes atomic displacements, and consequently lattice defect accumulation (e.g. point defects, dislocation loops, voids), inducing embrittlement and hardening [7]. Recent tungsten studies have shown that grain refinement reduces irradiation damage retention [8,9], as grain boundaries act as defect sinks. Nanocrystalline TiNi alloys [10] and 304L stainless steel [11] have also demonstrated enhanced irradiation resistance, and defect denuded zones near grain boundaries have been observed in He-ion-irradiated nanocrystalline Fe [12]. It should be noted that sputter-deposited thin Fe films, not severely plastically deformed materials, were considered in this latter study. Therefore, an interesting question is whether this approach also works in steels that are microstructurally complex and nanostructured using post processing/deformation methods.

Severe plastic deformation is an efficient means of inducing grain refinement in crystalline materials [13,14]. Interestingly, with severe plastic deformation, a saturated state with a stable nanocrystalline microstructure can be reached that shows little further evolution with continued deformation [15]. High-pressure torsion (HPT) involves the compression and subsequent shearing of samples, and provides a convenient way of accessing a diverse range of strain conditions in a single sample under controlled conditions [16,17].

In this study, we investigate the effects of severe plastic deformation, via HPT, on the irradiation hardening of Eurofer-97 as a prototypical ferritic/martensitic steel. We hypothesise that dense grain boundary networks will act as irradiation defect sinks, thus reducing irradiation-induced hardening.

## 2. **Experiments**

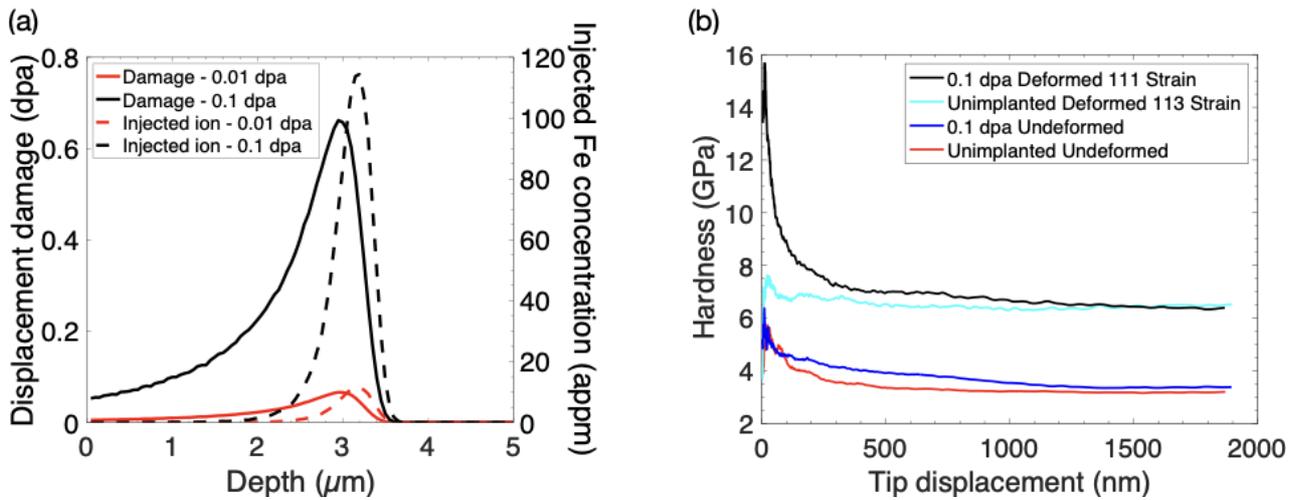

Figure 1. (a) Displacement damage vs. depth into the sample. Nominal sample irradiation levels were calculated from the average displacement damage up to 2 μm. Damage profile was calculated using Stopping and Range of Ions in Matter (SRIM) code [18] (Quick K-P calculation, 20 MeV Fe ions, pure Fe target with 40 eV displacement energy [19]) (b) Representative indentation hardness versus tip displacement curves.

Three Eurofer-97 (Heat number: 993394. Composition shown in table 1) disks with 5 mm diameter and 1 mm thickness were deformed by HPT [15]. The HPT process firstly applied 80 kN normal load corresponding to 4.1 GPa. Once loaded, samples were subjected to 9 turns at room temperature. HPT produces linearly increasing strains with radius, with shear strains ranging from 0 at the sample centre, to ~230 at the sample edge. Three samples were left undeformed. All samples were then mechanically polished with SiC paper, diamond suspension and colloidal silica. The final step of electropolishing was carried out with 5% perchloric acid in ethanol, with 15% ethylene glycol monobutyl ether, at 20°C using a voltage of 30 V for 2-3 minutes.

Table 1: Composition of the Eurofer-97 material [20].

| Element | C | Cr | Mn | V | N | W | Ta | Si |
|---|---|---|---|---|---|---|---|---|
| Wt% | 0.105 | 9.08 | 0.56 | 0.235 | 0.039 | 1.07 | 0.125 | 0.024 |

Implantation used 20 MeV $Fe^{3+}$ ions at the Helsinki Accelerator Laboratory (room temperature, 8 × $10^{-7}$ mbar vacuum). The ~5 mm diameter ion beam was rastered to achieve uniform ion fluences of 5.3 × $10^{13}$ cm$^{-2}$ and 5.3 × $10^{14}$ cm$^{-2}$, respectively corresponding to average nominal doses of 0.01 dpa and 0.1 dpa in the first 2 μm below the surface Fig. 1(a). Fluences were measured using a Faraday Cup in-front of and behind the sample. This depth range was chosen to calculate nominal dose as it is before the damage peak at the implantation range end. Implantation was carried out at a dose rate of 8.2 × $10^{-6}$ dpa/s, and an implantation duration of 201 minutes for 0.1 dpa, and 20.1 minutes for 0.01 dpa. The maximum temperature reached in the samples during the implantation was less than 50°C.

Nanoindentation was performed using an MTS Nano Indenter XP and a Berkovich tip, with the tip area function calibrated on fused silica. Continuous stiffness measurements (CSM) were made to 2 μm below the surface using a 0.05 s$^{-1}$ strain rate, 45 Hz CSM frequency, and a 2 nm harmonic amplitude. Samples deformed by HPT, due to radially increasing strains, were analysed using a row of indents across the sample diameter (50 μm spacing). Undeformed samples were analysed using a 5x5 indent grid (50 μm spacing). A transition from hardness dominated by the irradiated layer to a lower plateau representative of bulk hardness is seen at ~600 nm (Fig. 1(b)). At shallow indentation depths (< 200 nm), the indentation size effect dominates hardness, as described in the model by W.Nix and H.Gao [21]. To minimise size effects, while still probing the irradiated layer in implanted samples, average hardness over depth range 400 nm to 600 nm was extracted for all samples.

## 3. Results and discussion

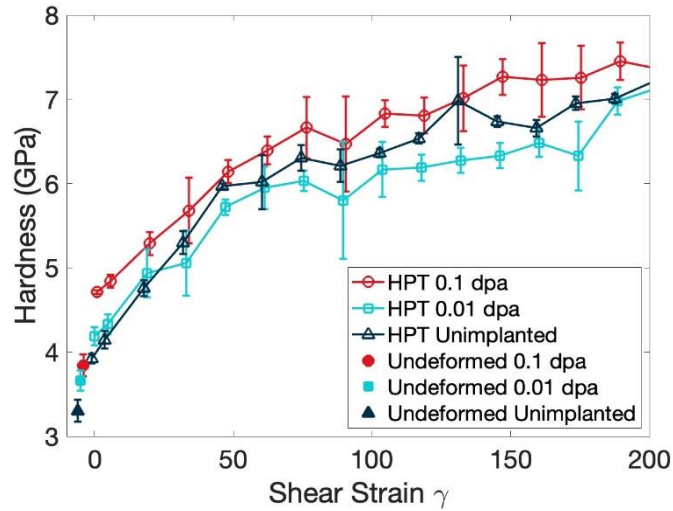

*Figure 2. Average hardness of unimplanted and implanted datasets plotted against shear strain ($\gamma$). Shaded areas represent ± one standard deviation of the indentation hardness. Undeformed data points have been offset by small negative shear strains for clarity.*

By converting each indent's radial position into shear strain ($\gamma$) [22] hardness variation with $\gamma$ and irradiation level can be plotted (Fig. 2). The hardness profile across the sample diameter is expected to be symmetric about the sample centre. This is due to the circular geometry of the samples, making the HPT-induced strain fields rotationally symmetric about the sample centre. This means that as a line of hardness measurements is taken across the diameter, for every pair of points either side of the sample centre, the strain should be nominally the same. Therefore, at each strain value, the hardness plotted is showing the average of datasets left and right of the centre point. Error bars show ± one standard deviation hardness value in the 400 – 600 nm range at each shear strain value. For undeformed samples (0 shear strain), average hardness values from 25 individual hardness measurements have been plotted. These values have been plotted at a small negative shear strains for clarity (Fig. 2).

The hardness of undeformed samples increases monotonically with irradiation damage due to irradiation hardening. The hardness change from unimplanted to 0.01 dpa is greater than from 0.01 dpa to 0.1 dpa in undeformed samples, suggesting hardness is approaching saturation, as previously reported for FeCr alloys [23,24]. Hardness also increases with strain due to strain hardening and grain refinement (Fig. 2). Grain refinement was seen experimentally via EBSD, where the initial area-weighted average grain diameter was 5.4 μm, and 140 nm after HPT (Appendix A). Furthermore, after $\gamma$ ~70, a reduction in the hardening with strain is shown which suggests that Eurofer-97 is approaching strain hardening saturation. Strain hardening "saturation" has been reported in literature e.g., for stainless steel samples processed via HPT [25].

Surprisingly, hardness at the centres of all the HPT samples, where $\gamma \sim 0$, is higher than in their undeformed counterparts (Fig. 2). This could be due to strain hardening associated with the initial sample compression before torsion is applied.

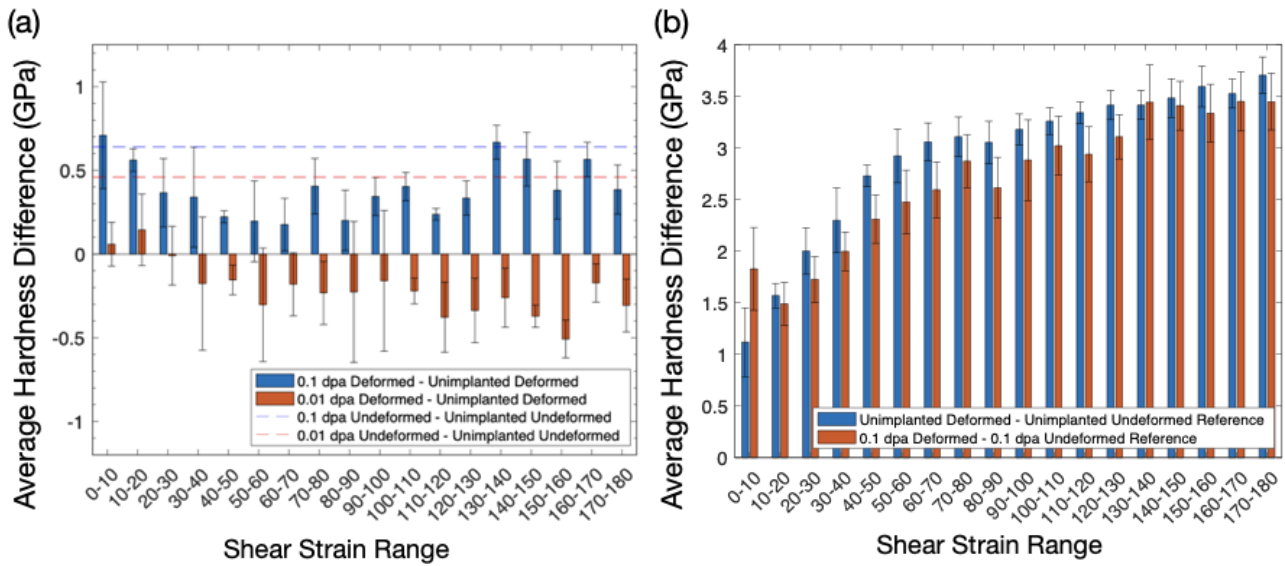

*Figure 3. (a) The irradiation-induced hardening for different doses with and without HPT processing prior to irradiation. (b) Comparison of average hardness difference between unimplanted HPT sample and the unimplanted undeformed material, and between the HPT samples implanted to 0.1 dpa and undeformed material implanted to 0.1 dpa.*

Fig. 3 (a) shows the relative change in hardness of irradiated HPT samples compared to their unirradiated counterpart as a function of shear strain. The average change in hardness is -0.21±0.16 GPa and 0.39±0.16 GPa for 0.01 dpa and 0.1 dpa respectively. Also shown is the average hardness increase of undeformed irradiated samples compared to undeformed unirradiated material, with 0.46±0.03 GPa for 0.01 dpa and 0.64±0.03 GPa for 0.1 dpa (dashed horizontal lines in Fig. 3 (a)). After irradiation to 0.1 dpa, HPT samples show relatively less irradiation hardening than the undeformed material (Fig.3 (b)). This may be due to grain boundaries acting as sinks for irradiation defects [26]. A smaller retained defect density means glide dislocations encounter fewer obstacles, resulting in less hardening. It is interesting to note that the 0.01 dpa HPT sample shows softening in the strain range ~60 to ~160. It has been proposed that this is due to irradiation-induced local thermal fluctuations unpinning glide dislocations, thus causing softening [27]. We hypothesise that both these mechanisms in fact contribute to the reduced irradiation hardening we observe in HPT samples. We also note that irradiation-induced hardening changes of HPT samples are small compared to deformation-induced hardening (Fig. 3(b)).

## 4. Conclusion

We have produced nano-structured Eurofer-97 by severe plastic deformation. Nanoindentation suggests that the microstructure is approaching saturation for shear strains greater than 70. Hardness increase due to ion-irradiation is smaller in severely deformed samples than in undeformed material. This suggests that nano-structuring by severe plastic deformation may provide a path to reducing irradiation hardening. This effect appears to be most prominent at low irradiation doses. We also note that hardening due to severe plastic deformation is much greater than that associated with irradiation.

All data supporting this study is available at: [Zenodo link to be added on acceptance]

## Acknowledgements:

We thank UKAEA for providing Eurofer-97 material and David Armstrong for nanoindenter use. GSP acknowledges funding from the Henry Royce Institute and the Oxford University Materials Science department, FH from ERC starting grant 714697, and KS from the General Sir John Monash Foundation.

## Appendix A – Grain maps of samples

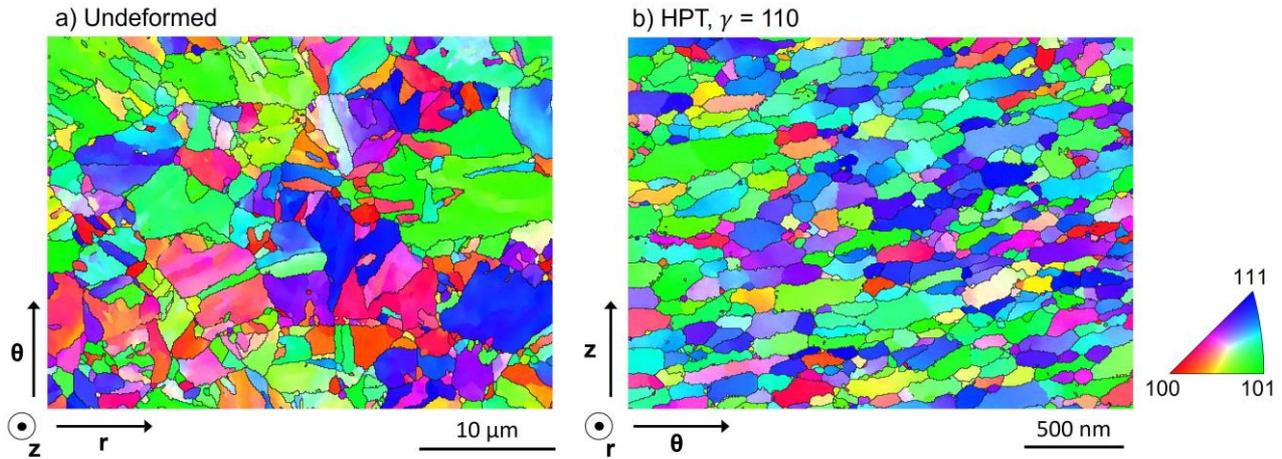

*Figure 4. Inverse pole figure grain maps of a) the undeformed and b) HPT sample ($\gamma$ =110). The sample coordinates and measurement methods are explained in the accompanying text. The black lines indicate grain boundaries with misorientation of >10°.*

The grain maps were measured with Zeiss Merlin FEG-SEM. The sample coordinates are defined in cylindrical coordinates with respect to the HPT sample where **z** is normal to the surface of the bulk sample, **θ** is the shear direction, and **r** is the radial direction.

For the undeformed sample, the top surface was measured with electron backscatter diffraction (EBSD). The average grain size was found to be 5.3 ± 3.2 µm. For the deformed sample, EBSD did not provide enough spatial resolution and a cross-sectional lift-out sample had to be made then measured with transmission Kikuchi diffraction (TKD). The average grain size is 146 ± 82 µm.

Credit Author Statement:

**Gregory Strangward-Pryce:** Conceptualisation, Software, Formal Analysis, Investigation, Data Curation, Writing-Original Draft, Writing-Review & Editing, Visualisation. **Kay Song:** Conceptualisation, Supervision, Project Administration, Software, Formal Analysis, Methodology, Validation, Investigation, Writing-Original Draft, Writing-Review & Editing, Visualisation, Resources, Data Curation. **Felix Hofmann:** Funding acquisition, Project Administration, Conceptualisation, Supervision, Writing-Original Draft, Writing-Review & Editing, Data Curation. **Kenichiro Mizohata:** Resources.